\documentclass[12pt]{article}

\usepackage{graphicx}
\usepackage{amsmath}

\providecommand{\AmS}{{\protect\the\textfont2
\renewcommand{\thesection}{\Roman{section}}
  A\kern-.1667em\lower.5ex\hbox{M}\kern-.125emS}}

\begin{document}
\centerline{\bf $\theta$-Vacuum Systems Via Real Action 
Simulations}
\vskip 2 truecm
\centerline { V.~Azcoiti$^a$, G. Di Carlo$^b$, A. Galante$^{c,b}$ and V. Laliena$^a$}
\vskip 1 truecm
\centerline {\it $^a$ Departamento de F\'\i sica Te\'orica, Facultad 
de Ciencias, Universidad de Zaragoza,}
\centerline {\it 50009 Zaragoza (Spain).}
\vskip 0.15 truecm
\centerline {\it $^b$ Istituto Nazionale di Fisica Nucleare, 
Laboratori Nazionali del Gran Sasso,}
\centerline {\it  67010 Assergi (L'Aquila), (Italy). }
\vskip 0.15 truecm
\centerline {\it $^c$ Dipartimento di Fisica dell'Universit\`a 
di L'Aquila,  67100 L'Aquila, (Italy).}
\vskip 3 truecm

\centerline {ABSTRACT}
Inspired by the results of the Ising model within an imaginary external 
magnetic field, we introduce a transformation in quantum systems with 
a $\theta$-vacuum term that amounts to a rescaling 
of $z=\cos({\theta\over2})$. 
Making use of this transformation we are able to 
determine the order parameter as a function of $\theta$.
The approach is successfully tested in models with both broken
and unbroken $CP$ symmetry at $\theta=\pi$.  

\newpage

Quantum systems with complex euclidean actions have received an important 
amount of attention in recent time. Indeed there are two very relevant 
subjects in quantum field theory the analysis of which 
requires the use of complex actions:
the study of topological structures and the behavior 
of matter at high temperature and density.
Notwithstanding the big interest in these subjects, 
it is currently not possible to write an efficient code to perform 
numerical simulations of systems with complex actions. All efficient 
algorithms to simulate statistical systems are based in the use of a true 
probability distribution function as Boltzmann weight. This is the 
reason why all attempts to analyze systems with complex actions are 
based $(i)$ on 
the use of specific properties of the particular model under study 
\cite{STURN}, $(ii)$ uncontrolled analytical continuations, 
or $(iii)$ in a very delicate process of summing up 
all the contributions which appear in the definition of the partition 
function \cite{PRD}, \cite{JAPAN}, \cite{BURK}, \cite{NIS1}, \cite{NIS2}.

In a recent paper \cite{monos} we proposed a new approach to analyze systems 
with a $\theta$-vacuum term, tested it in two analytically solvable models, 
and applied to the study of $\textrm{CP}^3$ with a topological term in the 
strong 
coupling region. This approach is based on the use of a high accuracy 
method to measure 
the probability distribution function of the topological charge density at 
$\theta = 0$ from numerical simulations at imaginary $\theta$.
It is combined 
with a multiprecision algorithm which allows to sum up all the terms which 
appear in the partition function, that differ by many orders of 
magnitude, running from 1 to $e^{-V}$, where $V$ is the lattice 
volume. 

Even if the results reported in \cite{monos} are quite 
promising, the method is not free from systematics and it would be therefore 
important to have alternative approaches available in order to compare  
physical predictions.
This is indeed the purpose of present paper.

We should say from the beginning that 
contrary to the approach reported in \cite{monos} where no assumption on the 
phase structure of the system was made, 
\textit{here we will make the assumption that 
the system has no phase transition at real $\theta$ except at most at 
$\theta = \pi$.} This means that our proposal should be used in practice as 
complementary to other approaches and one should look for consistency 
between their results. 

The method is based in extrapolating a suitably defined function to the origin.
The function turns out to be very smooth in all the cases we have considered 
up to
now and this makes us confident on the whole procedure.
At the very end it is equivalent to an analytical continuation from imaginary
to real $\theta$. However it has not to be confused with analytical 
continuation
of the free energy density and/or the order parameter which are extremely
sensitive to statistical errors and to the choice of fitting functions and 
therefore with small or no practical utility.

To fix the ideas and to keep this paper selfcontained, let us write 
the general 
form of the partition function of a system with a topological $\theta$-term 
as a sum over all topological sectors of the partition functions for each 
sector weighted with the proper topological weight,
\begin{equation}
{\cal Z}_V(\theta)\;=\;\sum_n\,p_V(n)\,{\mathrm e}^{{\mathrm i}\theta n}\, ,
\label{partition}
\end{equation}
where $p_V(n)$ is, up to a normalization constant, 
the probability of the topological sector $n$ at $\theta=0$, and the sum 
runs over all integers $n$. Writing equation (\ref{partition}) one 
needs to assume that $CP$ symmetry is realized at $\theta=0$ since otherwise 
$Z_{V}(\theta)$ would be ill-defined \cite{NOS}.

In almost all practical 
cases the sum in (\ref{partition}) 
has a number of terms of order $V$ since the 
maximum value of the topological charge at finite volume is of this order.
The partition function (\ref{partition}) can also be written 
as a sum over the density of topological charge $x_n =n/V$, and setting 
$p_V(n)=\exp[-V f_V(x_n)]$ one gets:
\begin{equation}
{\cal Z}_V(\theta)\;=\;\sum_{x_n}\,e^{-V f_V(x_n)}\,
{\mathrm e}^{{\mathrm i}\theta V x_n} \, .
\label{pfdiscrete}
\end{equation}
Equation (\ref{pfdiscrete}) defines a $2\pi$ periodic function of
$\theta$. 
For models $CP$-invariant at $\theta=0$ 
$f_V(x_n)$ is an even function; this property allow us to write:
\begin{equation}
{\cal Z}_V(\theta)\;=\;\sum_{y_n}\,G_{V}(y_n)\,
\left( {\cos^{2}{\theta\over2}}\right)^{V y_n} \, 
\label{pfdiscreten}
\end{equation}
where $y_n$, which is non negative by construction, 
takes the same non negative 
values as $x_n$ and $G_{V}(y_{n})$ is a 
functional of $f_{V}(x_{n})$. The mean value $y(\cos{\theta\over{2}})$ of 
$y_n$ computed from (\ref{pfdiscreten}) is related to the mean value 
$x(\theta)$ of $i$ times the density of topological charge as follows:
\begin{equation}
x(\theta)\;=\;i y\left(\cos{\theta\over2}\right)\tan{\theta\over2}\, 
\label{xtheta}
\end{equation}
which shows explicitly the $2\pi$ periodicity. At imaginary $\theta$, 
$\theta=-ih$, the previous equation becomes
\begin{equation}
x(-ih)\;=\;y\left(\cosh{h\over2}\right)\tanh{h\over2}\, 
\label{xh}
\end{equation}
Furthermore, at real $h$, the system described by (\ref{partition}) can 
easily be simulated by standard numerical algorithms since in this case the 
Boltzmann weight is well defined. From numerical simulations at real $h$ one 
can get $y(z)$ (\ref{xh}) for values of the argument 
$z=\cosh{h\over2}$ between 
1 and $\infty$. In order to get the topological charge 
density at real $\theta$ 
one should extend this function to the $(0,1)$ interval in $z$. But as stated 
before, an extension of $y(z)$ is sensitive to statistical 
errors and fitting functions so that it makes this procedure not reliable.

The function $y(z)$ depends also on the couplings of the model under 
consideration. For the one-dimensional Ising model within an external real or 
imaginary magnetic field $h$, the function $y(z)$ can be analytically 
computed and the final result is
\begin{equation}
y(z)\;=\;{{\left(e^{-4F}-1\right)^{-1/2} z}\over{\left(1+\left(e^{-4F}
-1\right)^{-1}z^2\right)^{1/2}}}\, 
\label{yz}
\end{equation}
which shows explicitly how this model, at imaginary values of the 
magnetic field $(0<z<1)$ and antiferromagnetic couplings 
$\left(F={J\over{KT}}<0\right)$ breaks spontaneously the $Z(2)$ symmetry at 
$h= i\pi$ (see (\ref{xtheta})). But this model shows also a very 
interesting property: a change in the coupling $F$, or what is the same, a 
change in the physical temperature $T$, is equivalent to a rescaling of the 
variable $z$ since $y$ depends on $z$ through the combination 
$e^{\lambda\over2}z$, with $\lambda = -\log(e^{-4F}-1)$. In other words, if 
we define the transformation:
\begin{equation}
y_{\lambda}(z)\;=\;y\left(e^{\lambda\over2}z\right)\, 
\label{ylambda}
\end{equation}
which would allow us, for negative values of $\lambda$, to extend the order 
parameter from imaginary $\theta$ to real $\theta$, the $1-d$ 
Ising model shows the 
exceptional feature that the transformation (\ref{ylambda}) is equivalent 
to a change of the coupling. This property, if true for other 
models, would be extremely interesting since it would reduce the problem of 
complex actions to simulations with real actions. Unfortunately this 
seems not to be the case since it is easy to check that already 
compact $U(1)$ in two 
dimensions with a $\theta$-term does not exhibit such a property. However 
this is not the end of the story and
we will try in what follows to make use of transformation (\ref{ylambda}) in 
a much less ambitious way.

Many of the physically relevant models, as $\textrm{CP}^{N-1}$ or $QCD$, are 
asymptotically free. Asymptotic freedom implies that the continuum limit is 
reached in the weak coupling region, but in this region the order parameter 
(and $y(z)$) will be very small for values of $z$ near 1 since 
the density of 
topological structures in lattice units is very suppressed at weak coupling. 
This means that,
from simulations at real $h$, we can plot $y_{\lambda}/y$ against $y$
up to values of $y$ very close to the 
origin, provided that $y(z)$ does not vanish for $z>0$.
We cannot exclude this possibility completely, 
but it does not happen in any
of the exactly solvable models we know. Indeed, it is easy to show that
$y(z)\neq 0$ for $z>0$ if all the coefficients $G_V(y_n)$ entering the right 
hand side of equation~(\ref{pfdiscreten}) are positive, as it happens in the 
models
studied in the present work. In such a case, the fluctuation-dissipation
theorem ensures that $y(z)$ is a non decreasing function of $z$.
Furthermore, if the order parameter does not diverge at 
$\theta=\pi$, $y(z)$ should vanish at least linearly with $z$ at $z=0$ 
(that corresponds to $\theta=\pi$).

Moreover, if
$y_{\lambda}/y$ as a function of $y$ shows a smooth
behavior near $y=0$, we can expect a simple extrapolation 
to $y=0$ being reliable. 
This point is crucial in what follows and obviously can not be 
taken for granted.
Apart from being reasonable it has been verified 
in all the models whose analytical solution is known to us.
In particular it is true for the one-dimensional Ising model, compact $U(1)$ 
in two dimensions, $\textrm{CP}^3$, and a toy model that exhibits no 
phase transition at
$\theta=\pi$, thus covering all the possible different 
qualitative possibilities.

In the Ising case we have done simulations of the model within an external  
real magnetic field $h$ at $F=-2.0$. This coupling was chosen in order to 
be in the ''weak coupling region'' where $y(z)$ is small near $z=1$. The 
model breaks spontaneously the $Z(2)$ symmetry at $\theta=-i h =\pi$ and 
therefore $y(z)$ vanishes linearly with $z$ (see (\ref{xtheta}), (\ref{xh})). 

If we define an effective exponent $\gamma_{\lambda}$ by the following 
equation:
\begin{equation}
\gamma_{\lambda}\;=\;{2\over\lambda} \log \left({{y_{\lambda}}\over
{y}}\right)\, 
\label{exponent}
\end{equation}
this exponent, in the limit $y(z)\rightarrow 0$ ($z\rightarrow 0$), will give 
the dominant power of $y(z)$ as a function of $z$ near $z=0$. A value for this 
exponent that goes to 1 when $y\rightarrow 0$ will imply spontaneous 
symmetry breaking, and
values between 1 and 2 will signal a second order phase 
transition at $\theta=\pi$ with a divergent susceptibility.
The reason is that the order parameter, 
$x=\tan(\theta/2) y[cos(\theta/2)]$, will behave as 
$(\pi-\theta)^{\gamma_\lambda-1}$ for $\theta\rightarrow\pi$. 

Fig. 1 reports $y_{\lambda}/y$ against $y$ for the Ising model at 
$\lambda=0.5$. The solid line is a quadratic fit of the points. This figure 
contains also the results for the effective exponent 
$\gamma_{\lambda}$ against 
$y$ (lower points). A simple quadratic fit predicts 
$\gamma_{\lambda}(0)= 0.997(3)$ i.e., 
spontaneous symmetry breaking. We have checked that, as expected, 
we get compatible results
using different values of $\lambda=O(1)$, 
and the same is true for the rest
of the models considered. 

The fit of $y_{\lambda}/y$ against $y$ 
together with the dependence of $y(z)$ on $z$ for $z\ge 1$
(computed from numerical simulations at real $h$) 
allows one to reconstruct the order parameter 
(\ref{xtheta}) at real $\theta$. To simplify the discussion let us assume 
we have computed $y(z_0)=y^0$. Next using 
the fit of Fig. 1 we plot $y_{\lambda}$ against $y$ and look in this plot 
for the value of $y=y^1$ which corresponds to $y_{\lambda}=y^0$ (see Fig. 2). 
It is simple to verify that $y^{1}=y(e^{-\lambda/2}z_0)$. The next 
step is to iterate this 
procedure: we replace $y^0$ by $y^1$ and get 
$y^{2}=y(e^{-\lambda}z_0)$, and 
so on. Once we have computed in this way $y(z)$ for as many values of 
$z\le 1$ as desired, the order parameter can be computed using equation 
(\ref{xtheta}).

Fig. 3 contains the results for the Ising order parameter. We report also 
in this figure the exact analytical 
results in order to put in evidence the agreement.

For two-dimensional compact $U(1)$ with topological charge we chose 
to work at
$\beta=1.0$ and $\lambda=0.5$. Figures similar to 1 and 3 were obtained for 
this model. The agreement between numerical and analytical results for 
the order parameter was even better than for the Ising model. 
Because of space limitations the corresponding figures are not
included.

The next model we want to analyze here is $\textrm{CP}^3$ with 
topological charge. 
Contrary to the Ising and two-dimensional compact $U(1)$ cases, in which we 
can compare with the exact analytical results, $\textrm{CP}^{N-1}$ models, 
which share 
many interesting properties with $SU(N)$, are not analytically solvable. 

We have adopted for
the action the ``auxiliary U(1) field'' formulation:
\begin{equation}
S_g = N\beta\sum_{n,\mu}(\bar{z}_{n+\mu}z_n U_{n,\mu} +
\bar{z}_n z_{n+\mu} \bar{U}_{n,\mu}-2)
\label{action}
\end{equation}
where $z_n$ is a $N$-component complex scalar field that satisfies
$\bar{z}_n z_n = 1$ and $U_{n,\mu}$is a $U(1)$ ``gauge field''.
The topological charge operator can be constructed 
directly from the $U(1)$ field:
\begin{equation}
S_\theta = 
i \frac{\theta}{2\pi}\sum_p\log(U_p) 
\label{topocharge}
\end{equation}
where $U_p$ is the product of the $U(1)$ field around the
plaquette and $-\pi < \log(U_p) \le \pi$.

Simulations were done at $\beta=0.4$. This value of the coupling is not  
in the weak-coupling scaling region, but the values of $y(z)$ near $z=1$ 
are smooth enough to allow a simple extension of $y_{\lambda}/y$ until 
$y=0$. Fig. 4 contains the effective exponent $\gamma_{\lambda}$ against 
$y$ at $\lambda=0.5$. A simple quadratic fit predicts $\gamma(0) = 1.03(5)$ 
i.e., spontaneous symmetry breaking at $\theta=\pi$. By imposing, as 
suggested by the previous fit, $\gamma(0) = 1.0$ in the fit of 
$y_{\lambda}/y$ against $y$, we get for the order parameter the results 
reported in Fig. 5. The continuous line in this figure 
represents the numerical results obtained within the approach 
reported in \cite{monos}. The very good agreement 
between the two approaches makes quite reliable the numerical results and 
$CP$ structure at $\theta=\pi$ of Fig. 5.

The models previously analyzed have a common feature: $CP$ symmetry is 
always spontaneously broken at $\theta=\pi$. In order to check that our 
approach works also in the symmetric case we have analyzed a simple 
toy model, with only one effective degree of freedom, where $CP$ 
symmetry is enforced at $\theta=\pi$. This model, which resembles very much 
the free instanton gas model, is described at finite volume $V$ by the 
following partition function:
\begin{equation}
{\cal Z}_V(\theta)\;=\;\left( 1+A \cos\theta\right)^V \, .
\end{equation}
The order parameter at real $\theta$ 
\begin{equation}
x(\theta)\;=\;i {{A \sin\theta}\over{1+A \cos\theta}}\, 
\end{equation}
shows explicitly $CP$ symmetry at $\theta=\pi$, as it should be since this 
is a model of one degree of freedom. We have not done for obvious reasons 
numerical simulations of this model at real $h=-i\theta$ but 
in order to work in realistic conditions,  
we modified the exact value of the 
order parameter at real $h$ by adding to it a random 
gaussian error of 1 percent.
Fig. 6 contains the exact analytical 
results for the order parameter at $A=0.005$ and the points obtained within 
our approach, using always for the parameter $\lambda$ defined in 
(\ref{ylambda}) the 
value 0.5. The agreement is quite good.

A final comment on the possibility of applying this approach to the inverse 
problem i.e., finite density $QCD$, is worthwhile. $QCD$ at imaginary 
chemical potential and staggered fermions can be simulated numerically 
because the fermion determinant is positive definite. Its partition 
function belongs to the general class 
of partition functions described by (\ref{pfdiscrete}), where $x_n$ now stands 
for the density of baryonic charge and $\exp[-V f_V(x_n)]$ 
is the probability distribution 
function of the baryon density at vanishing chemical potential. 
From numerical simulation of QCD at imaginary chemical potential 
one can measure the mean value of the density of baryonic charge and from it 
we can get $y(z)$ for $z$-values between 0 and 1.
In order to get the 
density of baryonic charge at real chemical potential we should extend 
$y(z)$ to $z$-values larger than 1. Summarizing we know $y_{\lambda}/y$ as 
a function of $y$ around $y=0$ and we need to extend this quantity to 
larger values of $y$. But since a phase transition is expected in QCD 
at finite chemical potential, a singular behavior of $y_{\lambda}/y$ 
as function of $y$ should also be expected, and any analytical extension of 
this quantity should fail when trying to reproduce the correct behavior.

This work has been partially supported by an INFN-CICyT collaboration and 
MCYT (Spain), grant FPA2000-1252. 
Victor Laliena has been supported by Ministerio de Ciencia y Tecnolog\'{\i}a 
(Spain) under the Ram\'on y Cajal program.

\newpage

\begin{figure}[!t]\
\centerline{\includegraphics*[width=5in,angle=90]{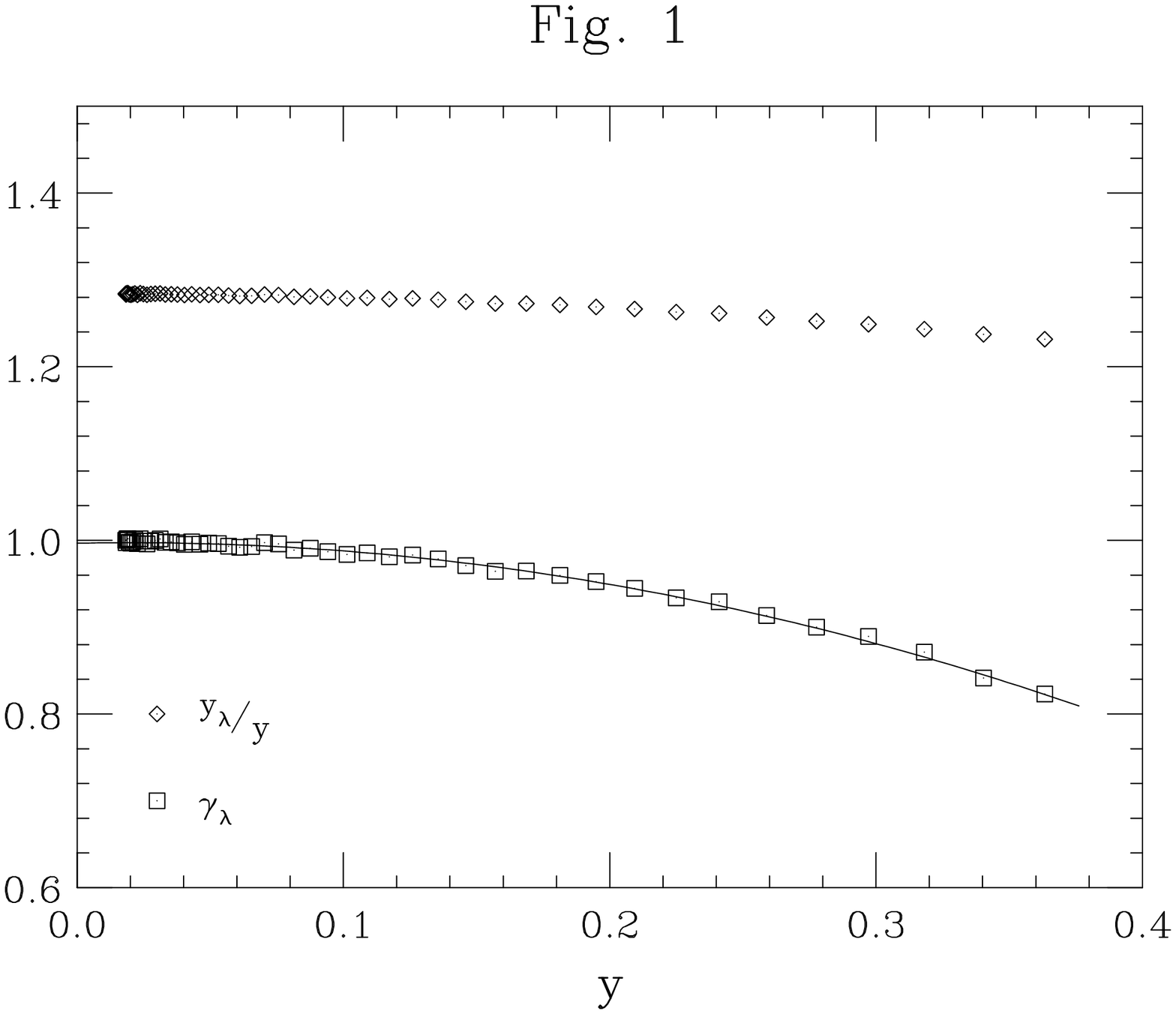}}
\caption{$y_\lambda/y$ and $\gamma_\lambda$ as a function of $y$ for
$\lambda=0.5$ in the Ising model ($F=-2.0$).}
\end{figure}

\begin{figure}[!t]\
\centerline{\includegraphics*[width=5in,angle=90]{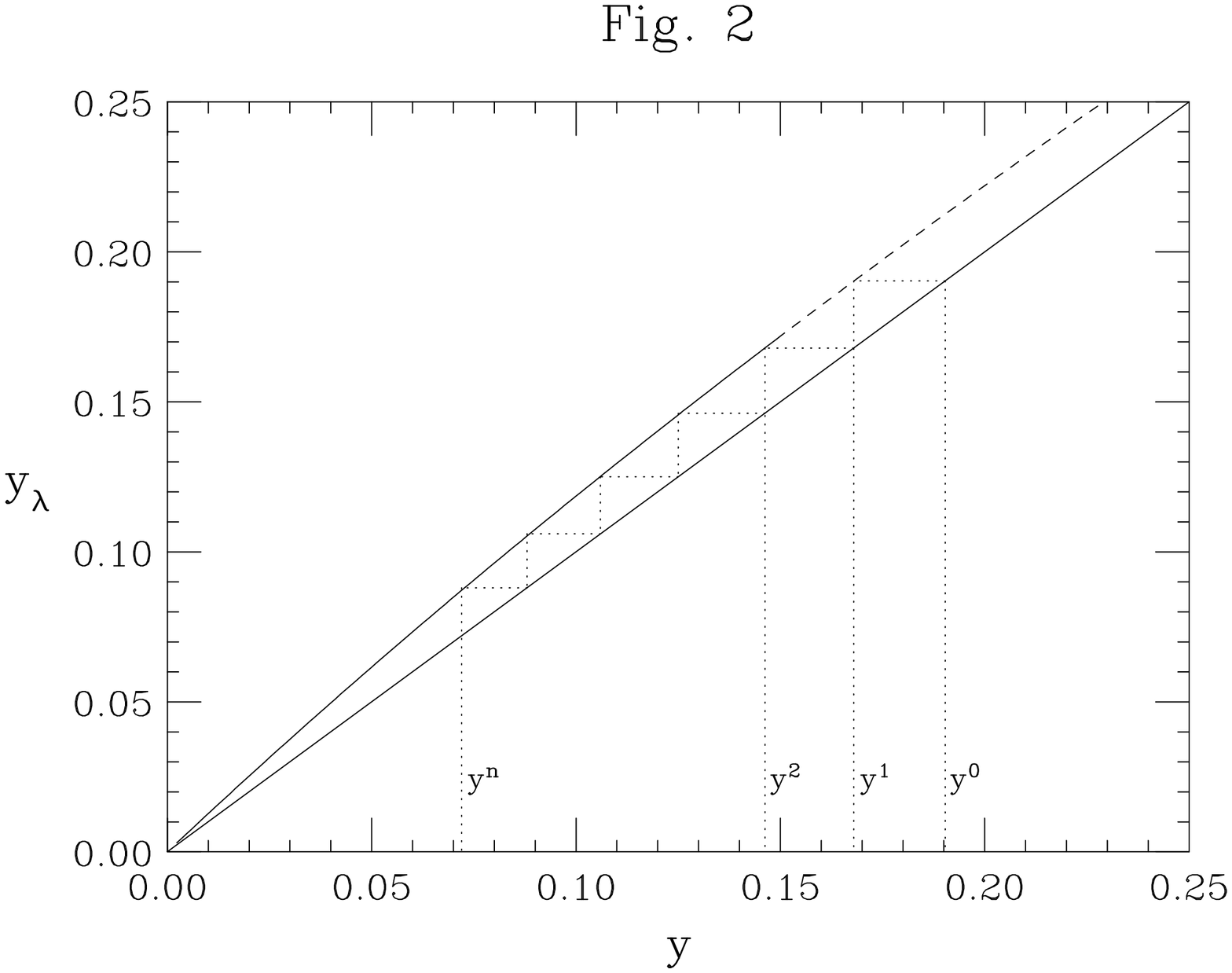}}
\caption{An example of the iterative procedure to reconstruct $y(z)$:
$y_\lambda$ as function of $y$ is plotted using dashed line in the region where
direct measurements are available and continuous line in the 
extrapolated region. The straight continuous line is $y_\lambda=y$.}
\end{figure} 

\begin{figure}[!t]\
\centerline{\includegraphics*[width=5in,angle=90]{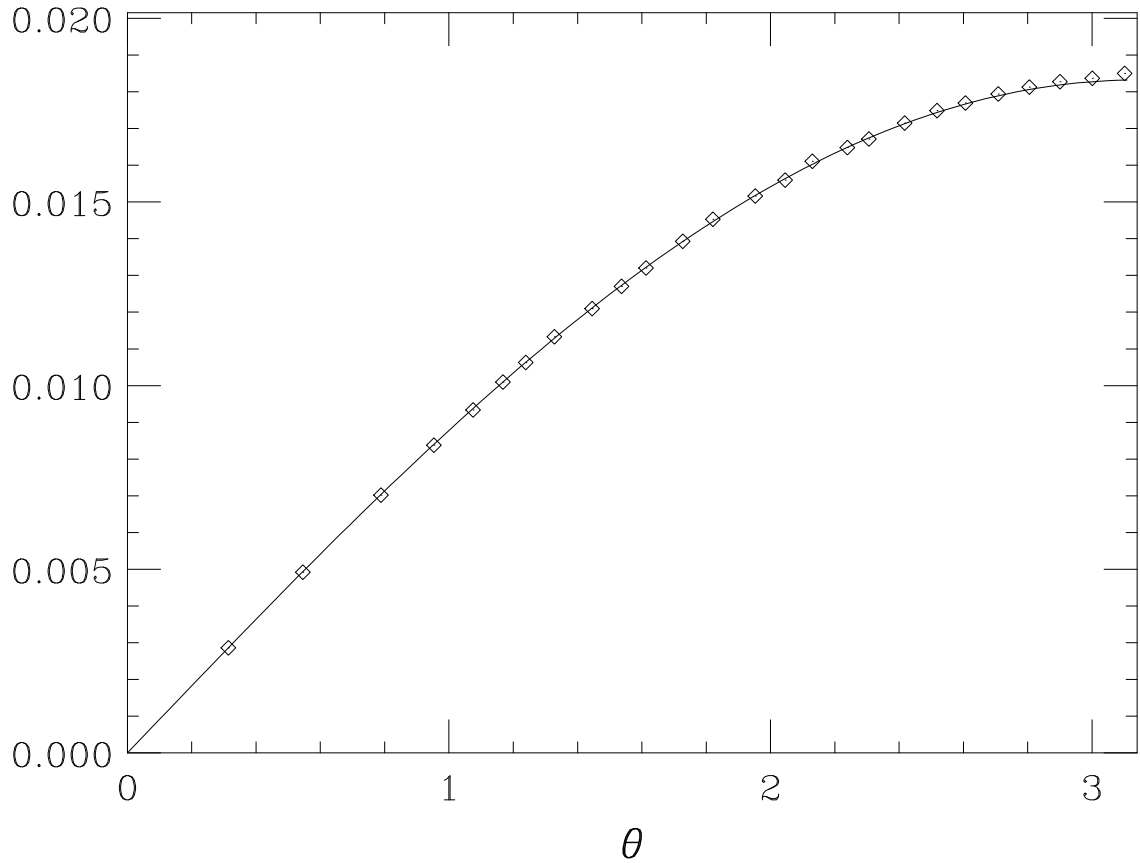}}
\caption{The order parameter in the Ising model ($F=-2.0$). Continuous line
is the exact result, points have been obtained from quadratic 
fit of $y_\lambda/y$ data reported in Fig. 1.}
\end{figure}

\begin{figure}[!t]\
\centerline{\includegraphics*[width=5in,angle=90]{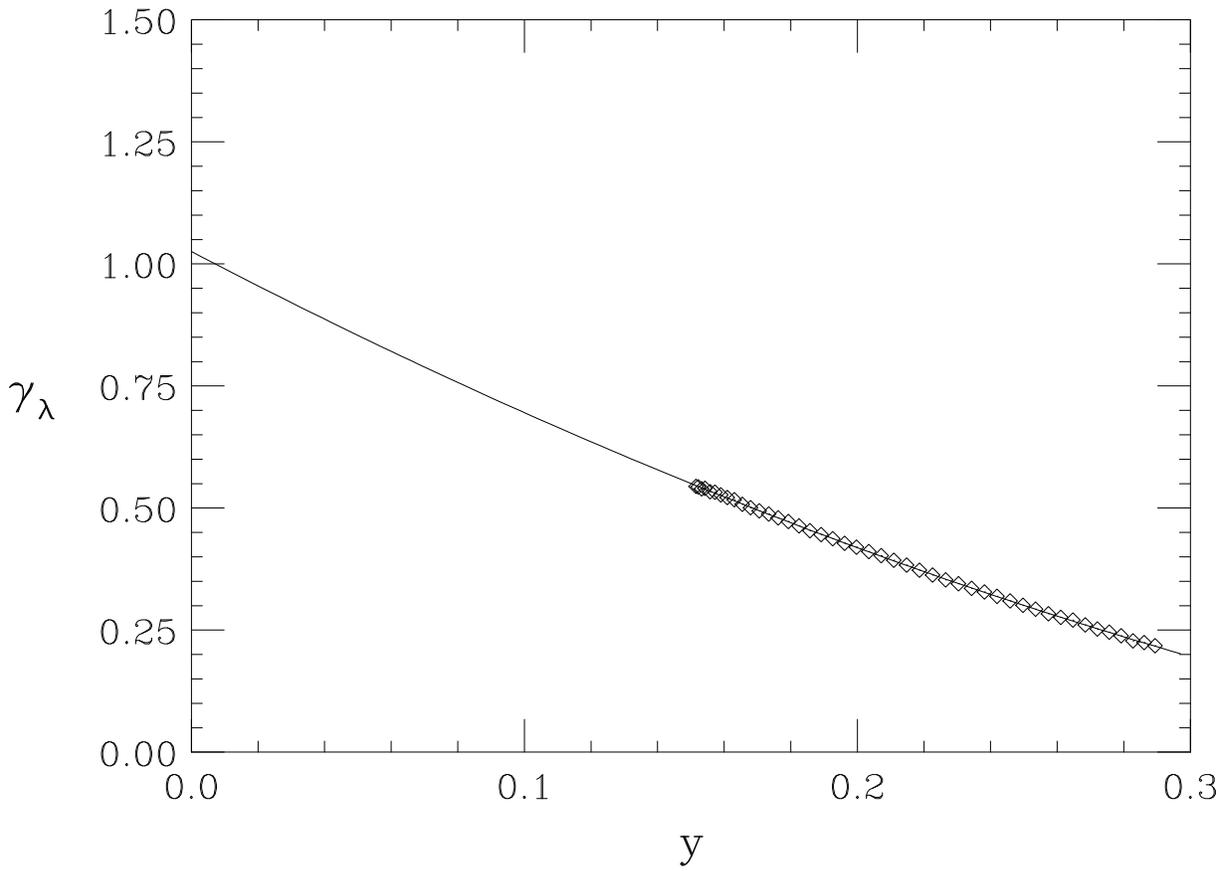}}
\caption{$\textrm{CP}^3$ model at $\beta=0.4$ in a 100x100 lattice: the 
effective exponent $\gamma_\lambda$ versus $y$.
The continuous curve is a quadratic fit to the data.}
\end{figure} 

\begin{figure}[!t]\
\centerline{\includegraphics*[width=5in,angle=90]{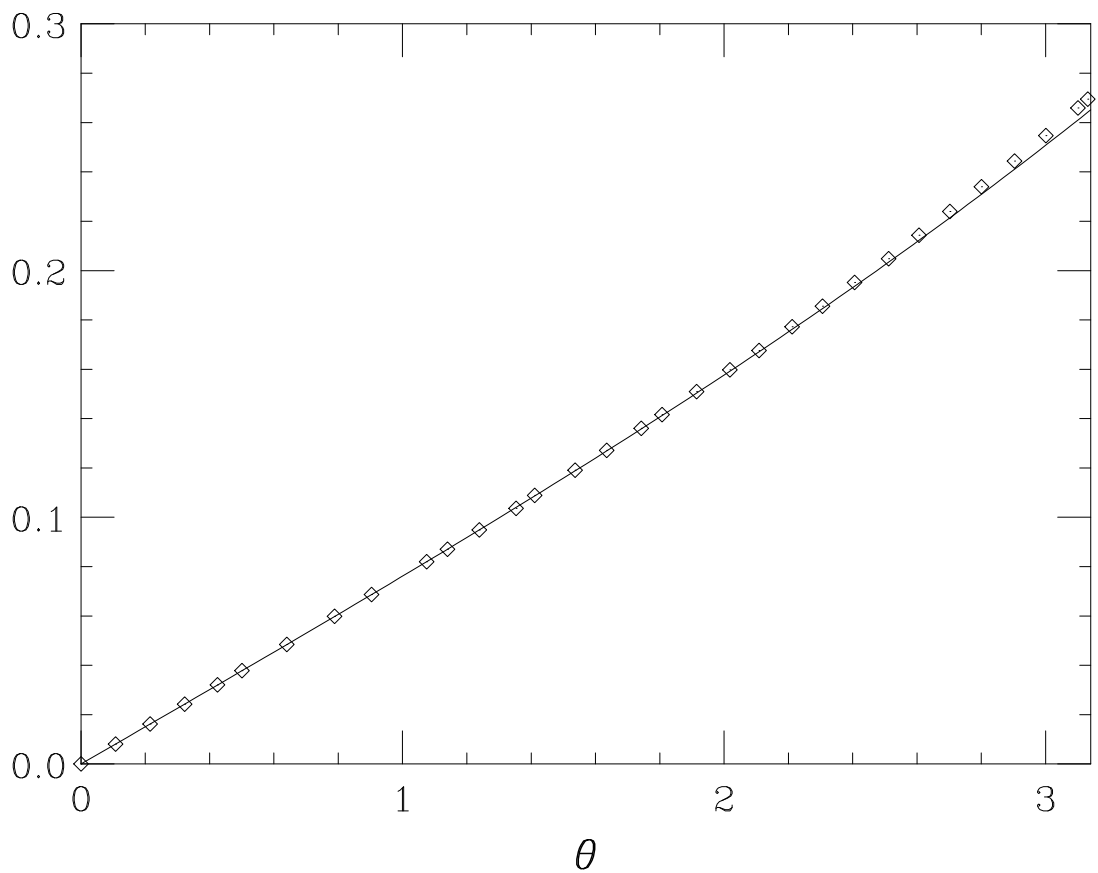}}
\caption{$\textrm{CP}^3$ model at $\beta=0.4$ in a 100x100 lattice. The 
order parameter as a function of $\theta$ from the fit of Fig. 4
(points) and the same quantity using the algorithm proposed in
\cite{NOS}.}
\end{figure} 

\begin{figure}[!t]\
\centerline{\includegraphics*[width=5in,angle=90]{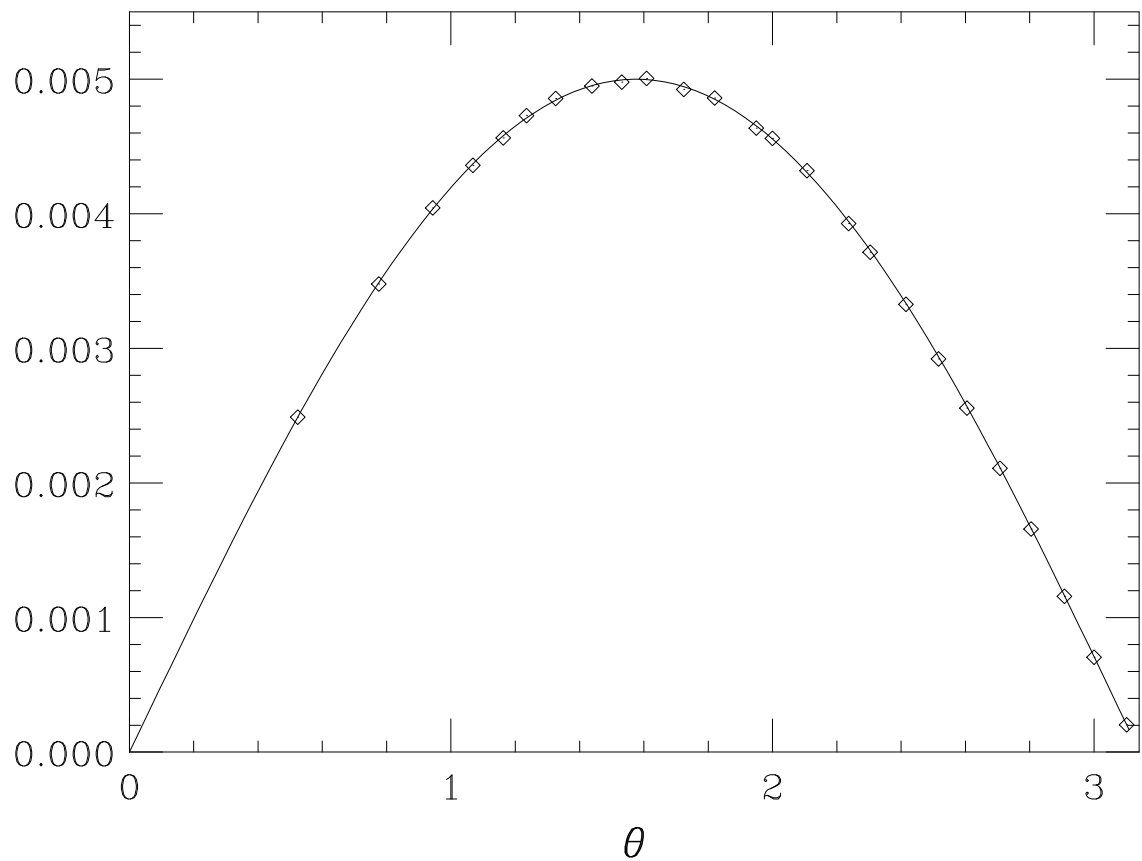}}
\caption{The $\textrm{CP}$ symmetric model presented in the text: 
numerical (points) and exact (line) results.}
\end{figure}

\end{document}